\documentclass[onecollarge,final]{svjour2}       
\smartqed  
\usepackage{graphicx,color}
\usepackage{mathptmx}      
\usepackage{amsmath,amssymb}
\newcommand{\re}{\mathrm{e}}

\newcommand{\unit}[1]{\,\mathrm{#1}}
\renewcommand{\(}{\left(}
\renewcommand{\)}{\right)}

\newcommand{\rd}{\mathrm{d}}

\newcommand{\expct}[1]{\langle #1 \rangle}

\newcommand{\tauf}{\tau_{\rm{}f}}
\newcommand{\taub}{\tau_{\rm{}b}}
\DeclareMathOperator{\sign}{sign}  

\renewcommand{\eqref}[1]{Eq.~(\ref{#1})}
\newcommand{\eqsref}[1]{Eqs.~(\ref{#1})}
\newcommand{\pref}[1]{(\ref{#1})}
\newcommand{\figref}[1]{Fig.~\ref{#1}}
\newcommand{\figsref}[1]{Figs.~\ref{#1}}
\begin{document}

\title{Characteristic Sign Renewals of Kardar-Parisi-Zhang Fluctuations}

\titlerunning{Characteristic Sign Renewals of KPZ Fluctuations}        

\author{Kazumasa A. Takeuchi         \and
        Takuma Akimoto 
}


\institute{
K. A. Takeuchi \at Department of Physics, Tokyo Institute of Technology, 2-12-1 Ookayama, Meguro-ku, Tokyo 152-8551, Japan. \\ \email{kat@kaztake.org}
 \and 
T. Akimoto \at Department of Mechanical Engineering, Keio University, Yokohama 223-8522, Japan
}

\date{Received: date / Accepted: date}

\maketitle

\begin{abstract}
Tracking the \textit{sign} of fluctuations governed
 by the $(1+1)$-dimensional Kardar-Parisi-Zhang (KPZ) universality class,
 we show, both experimentally and numerically,
 that its evolution has an unexpected link
 to a simple stochastic model called the renewal process,
 studied in the context of aging and ergodicity breaking.
Although KPZ and the renewal process are fundamentally different
 in many aspects, we find remarkable agreement
 in some of the time correlation properties,
 such as the recurrence time distributions and the persistence probability,
 while the two systems can be different in other properties.
Moreover, we find inequivalence between long-time and ensemble averages
 in the fraction of time occupied by a specific sign
 of the KPZ-class fluctuations. 
The distribution of its long-time average
 converges to nontrivial broad functions,
 which are found to differ significantly from that of the renewal process,
 but instead be characteristic of KPZ.
Thus, we obtain a new type of ergodicity breaking
 for such systems with many-body interactions.
Our analysis also detects qualitative differences
 in time-correlation properties of circular and flat KPZ-class interfaces,
 which were suggested from previous experiments and simulations
 but still remain theoretically unexplained.

\end{abstract}


\section{Introduction}

The Kardar-Parisi-Zhang (KPZ) universality class
 \cite{Kardar.etal-PRL1986,Barabasi.Stanley-Book1995,Kriecherbauer.Krug-JPA2010,Corwin-RMTA2012} is
 a prominent nonequilibrium class,
 ruling diverse kinds of nonlinear fluctuations in growing interfaces
 \cite{Kardar.etal-PRL1986,Barabasi.Stanley-Book1995,Takeuchi-JSM2014},
 driven particle systems \cite{Kriecherbauer.Krug-JPA2010,Corwin-RMTA2012},
 fluctuating hydrodynamics \cite{Spohn-a2015}, and so on.
Particularly noteworthy are recent analytical developments
 on the $(1+1)$-dimensional KPZ class,
 which have exactly determined a number of its statistical properties
 on the solid mathematical basis
 \cite{Kriecherbauer.Krug-JPA2010,Corwin-RMTA2012}.
Specifically, for $(1+1)$-dimensional KPZ-class interfaces,
 the interface height $h(x,t)$, measured along the growth direction
 at lateral position $x$ and time $t$, grows as
\begin{equation}
 h(x,t) \simeq v_\infty t + (\Gamma t)^{1/3} \chi(x,t)  \label{eq:Height}
\end{equation}
 with parameters $v_\infty$ and $\Gamma$,
 a rescaled random variable $\chi$,
 and $\beta \equiv 1/3$ being the characteristic growth exponent
 of the $(1+1)$-dimensional KPZ class
 \cite{Kardar.etal-PRL1986,Barabasi.Stanley-Book1995}.
Then the recent analytical studies
 \cite{Kriecherbauer.Krug-JPA2010,Corwin-RMTA2012} have consistently shown that
 $\chi(x,t)$ exhibits one of the few universal distribution functions,
 selected by the choice of the initial condition, or equivalently
 the global shape of the interfaces.
For example,
 circular interfaces grown from a point nucleus
 show the largest-eigenvalue distribution of random matrices
 in Gaussian unitary ensemble, called the GUE Tracy-Widom distribution,
 while flat interfaces on a linear substrate
 show the equivalent for Gaussian orthogonal ensemble.
This implies that the KPZ class splits into a few universality subclasses.
They are also characterized
 by different spatial correlation functions,
 whose exact forms are also known analytically
 \cite{Kriecherbauer.Krug-JPA2010,Corwin-RMTA2012}.
These results for the circular and flat subclasses were corroborated
 by direct experimental verifications using growing interfaces
 of turbulent liquid crystal (LC)
 \cite{Takeuchi.Sano-PRL2010,Takeuchi.etal-SR2011,Takeuchi.Sano-JSP2012}.

In contrast to these clear characterizations
 of the distribution and the spatial correlation,
 analytical results on the temporal correlation remain limited,
 hence challenging
 \cite{Dotsenko-JSM2013,Johansson-a2015,Dotsenko-a2015,Ferrari.Spohn-a2016}.
The LC experiment \cite{Takeuchi.Sano-JSP2012}
 and numerical simulations \cite{Kallabis.Krug-EL1999,Singha-JSM2005,Takeuchi-JSM2012,Carrasco.etal-NJP2014}
 showed that the temporal correlation
 is also different between the circular and flat subclasses.
Firstly, the two-time correlation function \footnote{
Throughout the paper, $\expct{\cdots}$ denotes the ensemble average
 defined over infinitely many realizations.
It is independent of $x$ because of the translational symmetry.
Therefore, for evaluation, we can take averages over positions too
 to achieve better statistical accuracy,
 without changing its mathematical definition.}
\begin{align}
 C(t, t_0)
 &\equiv \expct{h(x,t)h(x,t_0)} - \expct{h(x,t)}\expct{h(x,t_0)}  \notag \\
 &\simeq (\Gamma^2 t_0 t)^{1/3} C_{\rm res}(t/t_0)  \label{eq:CorrFunc}
\end{align}
 was shown to decay, in its rescaled form,
 as $C_{\rm{}res}(t/t_0)\sim(t/t_0)^{-\bar\lambda}$
 with $\bar\lambda=1$ for the flat case
 \cite{Takeuchi.Sano-JSP2012,Kallabis.Krug-EL1999,Carrasco.etal-NJP2014} and
 $\bar\lambda=1/3$ for the circular case \cite{Takeuchi.Sano-JSP2012,Singha-JSM2005,Takeuchi-JSM2012,Carrasco.etal-NJP2014}.
The latter implies
 $\lim_{t\to\infty}C(t,t_0)>0$, i.e.,
 correlation remains strictly positive,
 forever, in the circular case.
Secondly, the persistence probability $P_\pm(t,t_0)$
 was also measured, which is
 defined here by
 the probability that the fluctuation
 $\delta{}h(x,t)\equiv{}h(x,t)-\expct{h(x,t)}$ at a fixed position $x$
 never changes its sign (from the one denoted by the subscript)
 in the time interval $[t_0,t]$\footnote{
In the literature, our definition of $P_\pm(t,t_0)$
 based on the ensemble average
 is sometimes called the ``survival probability'',
 in which case the term ``persistence probability'' is reserved
 for the probability that the sign of $h(x,t)-h(x,t_0)$
 is unchanged \cite{Bray.etal-AP2013}.
However, in the present paper, we define our persistence probability
 by the ensemble average (unless otherwise stipulated),
 following earlier studies of direct relevance.}.
This quantity was found to show a power-law decay
\begin{equation}
 P_\pm(t,t_0) \sim (t/t_0)^{-\theta^{\rm(p)}_\pm}  \label{eq:PersProb}
\end{equation}
 with exponents
 $\theta_+^{\rm(p)}\approx1.35<\theta_-^{\rm(p)}\approx1.85$
 for the flat interfaces
 \cite{Kallabis.Krug-EL1999,Takeuchi.Sano-JSP2012}\footnote{
$\theta_+^{\rm(p)}<\theta_-^{\rm(p)}$ holds when the underlying
 KPZ nonlinearity $\frac{\lambda}{2}(\nabla{}h)^2$ is positive,
 otherwise the order is reversed \cite{Kallabis.Krug-EL1999}.
Note also that the estimates obtained in \cite{Kallabis.Krug-EL1999}
 are somewhat different from those
 from the LC experiment \cite{Takeuchi.Sano-JSP2012}.
We believe this is because of the different choice of the reference time
 $t_0$: while it was taken to be right after the initial condition
 in \cite{Kallabis.Krug-EL1999},
 times in the asymptotic KPZ regime were used
 in the LC experiment \cite{Takeuchi.Sano-JSP2012}.
This is also the choice of the present paper;
 therefore, here we refer to the estimates from the LC experiment,
 $\theta_+^{\rm(p)}\approx1.35$ and $\theta_-^{\rm(p)}\approx1.85$,
 as the values of the persistence exponents for the flat KPZ subclass.}
 and $\theta_+^{\rm(p)}\approx\theta_-^{\rm(p)}\approx0.8$
 for the circular ones
 \cite{Singha-JSM2005,Takeuchi.Sano-JSP2012,Takeuchi-JSM2012}.
The latter implies divergence of the mean persistence time
 $\int_{t_0}^\infty{}P_\pm(t,t_0)\rd{}t$ in the circular case.

It has been shown that such a divergent mean leads to anomalous dynamics
 such as non-ergodicity, anomalous diffusion, aging,
 and population splitting \cite{Godreche.Luck-JSP2001,Margolin.Barkai-JSP2006,Schulz.etal-PRL2013,Schulz.etal-PRX2014}.
Therefore, the above observations on the circular KPZ subclass
 imply that it may also be understood in this line of research.
Ergodicity is a basic concept
 in statistical physics and dynamical systems. 
It guarantees that
 time-averaged observables obtained by single trajectories converge 
 to a constant (ensemble average) as time goes on.
Ergodicity breaks down when, e.g., the phase space
 consists of mutually inaccessible regions,
 because then single trajectories are unable
 to cover the whole phase space.
However, in 1992, Bouchaud \cite{Bouchaud-JPIF1992}
 proposed another situation of ergodicity breaking,
 where the phase space is not split,
 but trajectories undergo long and random trapping.
If the mean trapping time diverges,
 trajectories cannot sufficiently explore the phase space,
 however long they do.
Interestingly, in this situation named weak ergodicity breaking (WEB)
 \cite{Bouchaud-JPIF1992}, certain time-averaged observables
 such as the time-averaged diffusion coefficient
 \cite{He.etal-PRL2008,Miyaguchi.Akimoto-PRE2011,Akimoto.Miyaguchi-PRE2013,Miyaguchi.Akimoto-PRE2015,Metzler.etal-PCCP2014}
 and the fraction of time occupied by a given state
 \cite{Lamperti-TAMS1958,Godreche.Luck-JSP2001,Margolin.Barkai-PRL2005,Margolin.Barkai-JSP2006} do not converge to their ensemble average,
 but themselves become well-defined random variables,
 described by characteristic distribution functions in simple cases.
The existence of such an asymptotic broad distribution
 for time-averaged quantities is usually regarded
 as a defining property of WEB.
Experimentally, single-particle observations have indeed shown
 relevance of WEB in macromolecule diffusion in biological systems
 \cite{Wong.etal-PRL2004,Jeon.etal-PRL2011,Weigel.etal-PNAS2011,Tabei.etal-PNAS2013,Manzo.etal-PRX2015,Metzler.etal-PCCP2014}
 and in blinking quantum dots \cite{Kuno.etal-JCP2000,Brokmann.etal-PRL2003,Stefani.etal-NJP2005,Stefani.etal-PT2009}.
However, it remains unclear,
 both theoretically and experimentally, how useful
 these developments on WEB are to characterize
 many-body problems such as the KPZ class.
Therefore, it is a challenging and important issue to clarify
 if WEB occurs in KPZ,
 and if yes,
 characterize the WEB of the KPZ class.

\section{Our approach}

To investigate possible relationship between WEB and KPZ,
 we construct a dichotomous process from height fluctuations
 of KPZ-class interfaces,
 $\sigma(x,t)\equiv\sign[\delta{}h(x,t)]$,
 which is then regarded as a time series.
Such dichotomization has recently been used to characterize
 time-correlation properties of interactions on lipid membranes
 \cite{Yamamoto.etal-SR2015,Yamamoto.etal-SR2015a}
 and of turbulence \cite{Herault.etal-EL2015,Herault.etal-JSP2015},
 successfully.
The constructed process is compared with
 a theoretically defined dichotomous process,
 arguably the simplest and best-studied one,
 namely the renewal process (RP) 
 \cite{Feller-Book1971,Cox-Book1962,Godreche.Luck-JSP2001}.
RP consists of a single
 two-state variable, which switches from one to the other state 
 after random, uncorrelated waiting times
 generated by a power-law distribution:
\begin{equation}
 p(\tau) \equiv \text{Prob}[\text{waiting time} > \tau] =\! \(\frac{\tau}{\tau_0}\)^{-\theta}\hspace{-12pt},  \hspace{9pt}(\tau \geq \tau_0). \label{eq:WaitingTimeRP}
\end{equation}
This model shows WEB and aging for $0<\theta<1$
 \cite{Feller-Book1971,Cox-Book1962,Godreche.Luck-JSP2001,Schulz.etal-PRX2014}.

Concerning KPZ-class interfaces,
 we use the experimental data
 of the circular and flat interfaces
 obtained in Refs.~\cite{Takeuchi.etal-SR2011,Takeuchi.Sano-JSP2012}
 (LC turbulence): for the circular (or flat) case,
 total observation time was $T_{\rm{}tot}=30.5\unit{s}~(63\unit{s})$,
 time resolution was $T_{\rm{}res}=0.5\unit{s}~(0.35\unit{s})$,
 and $N=955~(1128)$ realizations were used,
 respectively.
We also analyze newly obtained numerical data for circular interfaces
 of the off-lattice Eden model \cite{Takeuchi-JSM2012}
 ($T_{\rm{}tot}=5000,T_{\rm{}res}=1,N=5000$)
 and flat interfaces of the discrete polynuclear growth (dPNG) model
 ($T_{\rm{}tot}=10^4,T_{\rm{}res}=0.1,N=10^4$).
Further descriptions of the systems and parameters
 are given in \ref{app:a}.

\section{Results}

\subsection{Circular interfaces}

\begin{figure}[t!]
 \centering
 \includegraphics[width=.8\hsize,clip]{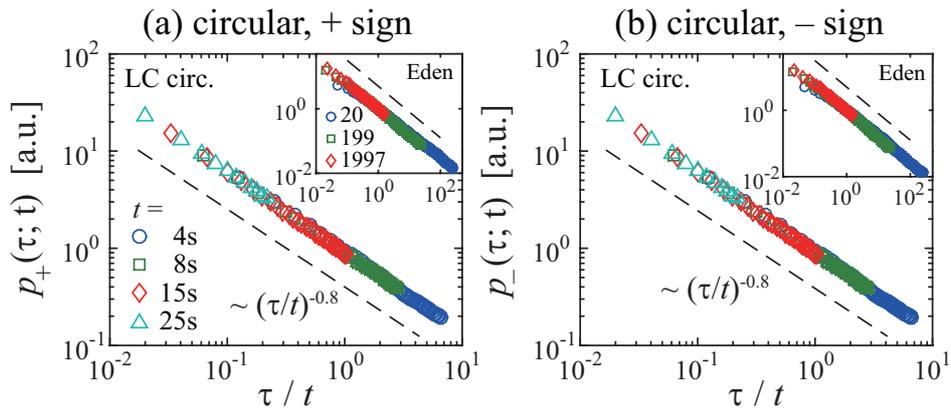}
 \caption{
Waiting-time distributions (ccdfs) $p_\pm(\tau;t)$ against $\tau/t$
 for the circular interfaces, obtained at different $t$
 in the LC experiment (main panels) and the Eden model (insets).
The dashed lines are guides to the eyes indicating exponent $-0.8$.
The same set of colors/symbols and $t$ is used in both panels.
The ordinates are arbitrarily shifted.
}
 \label{fig1a}
\end{figure}%

First of all, we stress that RP is far too simple to fully describe KPZ,
 because RP is a two-state model without even spatial degrees of freedom
 and has uncorrelated waiting times.
We nonetheless measure the waiting times between two sign changes
 of $\delta{}h$, first for the circular interfaces,
 for which we anticipate relation to WEB 
 as discussed above.
More specifically, we define the waiting-time distribution $p_\pm(\tau;t)$
 by the probability that the sign renewed at time $t$
 (changed to the subscripted one)
 lasts over time length $\tau$ or longer,
 hence $p_\pm(\tau;t)$ is the complementary
 cumulative distribution function (ccdf).
Figure~\ref{fig1a} shows the results
 for both the LC experiment (main panels) and the Eden model (insets).
Remarkably, in both cases we find a clear power law
 as described in \eqref{eq:WaitingTimeRP}
 with exponent $\theta=0.8$, while the cutoff $\tau_0$ is
 out of the range of our resolution.

\begin{figure}[t!]
 \centering
 \includegraphics[width=.8\hsize,clip]{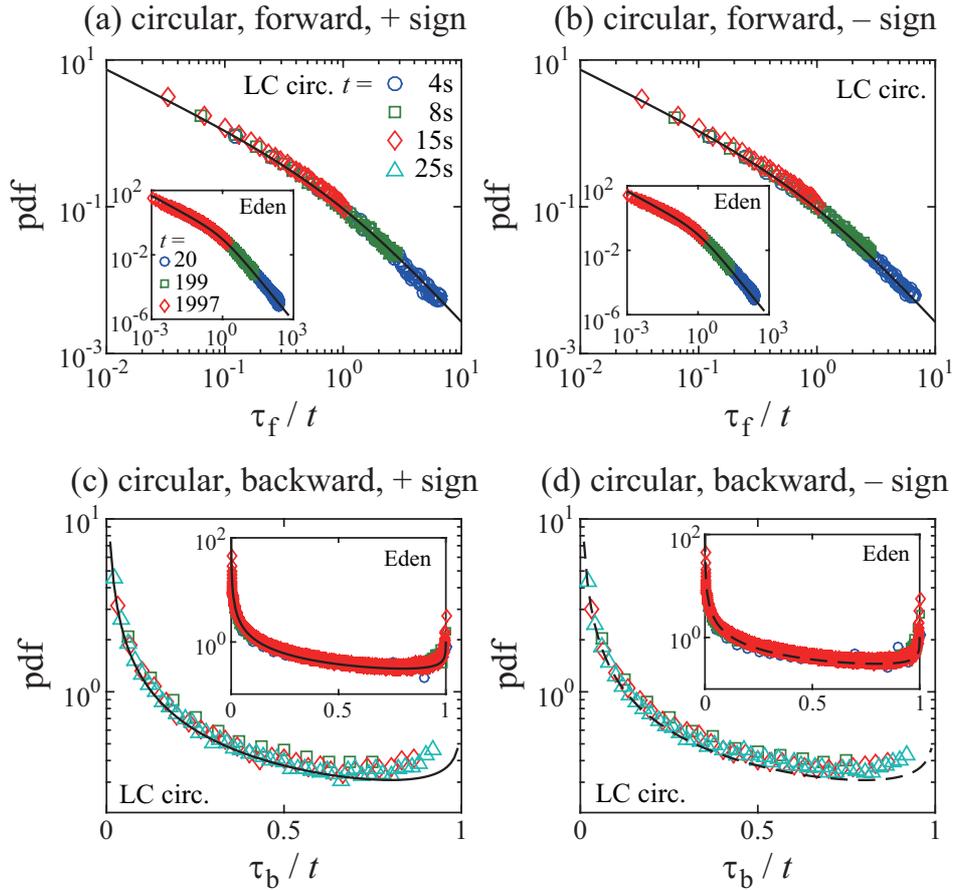}
 \caption{
Distributions (pdfs) of the forward (a,b) and backward (c,d) recurrence times,
 $\tauf$ and $\taub$, respectively,
 for the circular interfaces at different $t$
 in the LC experiment (main panels) and the Eden model (insets).
$\tauf$ and $\taub$ are rescaled by $t$.
The black lines indicate RP's exact results,
 \eqsref{eq:ForwRecurRP} and \pref{eq:BackRecurRP}, with $\theta=0.8$.
The data are normalized so that
 they have the same statistical weight as RP's exact results
 in the range covered by their abscissa.
The same set of colors/symbols and $t$ is used in all panels.
Note that data at $t=25\unit{s}$ are not shown for $\tauf$
 because the remaining time is then too short to measure
 the distribution of $\tauf$.
}
 \label{fig1b}
\end{figure}%

This similarity to RP leads us to
 compare further statistical properties between the two systems.
First we focus on the forward recurrence time $\tauf(t)$,
 defined as the interval between time $t$ and the next sign change,
 as well as the backward recurrence time $\taub(t)$,
 which is the backward interval from $t$ to the previous sign change
 \cite{Godreche.Luck-JSP2001,Dynkin-IANSSM1955,Feller-Book1971}.
For RP, Dynkin
 \cite{Dynkin-IANSSM1955,Godreche.Luck-JSP2001}
 derived exact forms
 of the probability density function (pdf)
 of $\tauf(t)$ and $\taub(t)$ as follows, for $0<\theta<1$:
\begin{align}
 &\text{pdf}(\tilde\tauf) = \frac{\sin\pi\theta}{\pi}\frac{1}{\tilde\tauf^\theta (1+\tilde\tauf)},  \label{eq:ForwRecurRP} \\
 &\text{pdf}(\tilde\taub) = \beta_{1-\theta,\theta}(\tilde\taub) \equiv \frac{\sin\pi\theta}{\pi}\tilde\taub^{-\theta}(1-\tilde\taub)^{\theta-1},  \label{eq:BackRecurRP}
\end{align}
 with $\tilde\tauf\equiv\tauf/t,\tilde\taub\equiv\taub/t$
 and $\beta_{a,b}(x)$ denoting the pdf of the beta distribution.
Although their derivation essentially relies
 on the independence of waiting times in RP,
 a feature not shared with KPZ,
 we find, as shown in \figref{fig1b},
 that both experimental and numerical results for the circular interfaces
 precisely follow RP's exact results indicated by the solid lines
 (except finite-time corrections).
Note that the persistence probability $P_\pm(t,t_0)$ considered
 in \eqref{eq:PersProb} actually
 amounts to the ccdf of $\tauf(t_0)$, i.e.,
 $P_\pm(t,t_0)=\int_{t-t_0}^\infty\text{pdf}(\tauf(t_0))\rd\tauf$.
This indicates that the functional form of the persistence probability,
 which is usually intractable
 for such spatially-extended nonlinear systems
 \cite{Bray.etal-AP2013},
 seems to be given by RP's exact result \pref{eq:ForwRecurRP}
 in the case of the circular KPZ subclass.
We also remark that
 the explicit dependence of the pdfs on $t$ indicates
 the aging of the system.

\begin{figure}[t!]
 \centering
 \includegraphics[width=.8\hsize,clip]{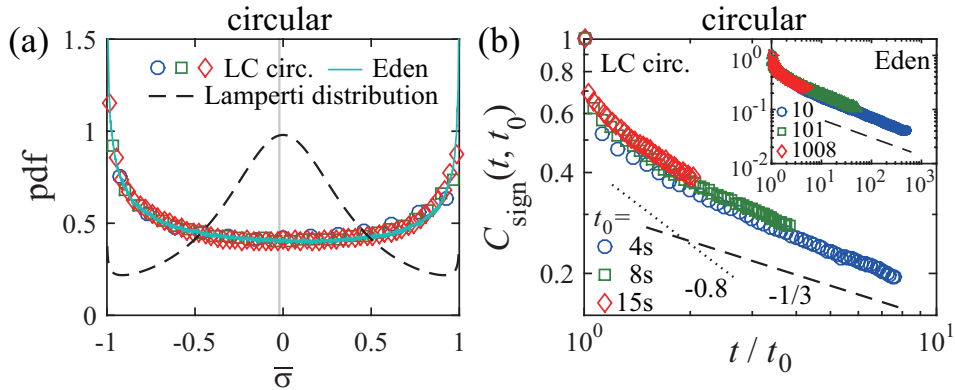}
 \caption{
Time-averaged sign distribution (a) and sign correlation function (b)
 for the circular interfaces.
(a) Pdf of the time-averaged sign
 $\bar\sigma=(1/T)\int_0^T\sigma(x,t)\rd{}t$
 for the LC experiment ($T=7.5\unit{s},15\unit{s},30.5\unit{s}$
 for circles, squares, and diamonds, respectively)
 and the Eden model ($T=5000$), compared to the Lamperti distribution
 \pref{eq:Lamperti} with $\theta=0.8$ (dashed line).
The gray vertical line indicates the ensemble-averaged value
 $\expct{\sigma}=-0.021$.
The existence of the broad asymptotic distribution is a direct evidence
 of WEB in the circular KPZ subclass.
(b) Correlation function of sign,
 $C_{\rm{}sign}(t,t_0)=\expct{\sigma(x,t)\sigma(x,t_0)}$,
 measured at different $t_0$
 for the LC experiment (main panel) and the Eden model (inset).
The dashed and dotted lines indicate
 exponents $-\bar\lambda=-1/3$ and $-\theta=-0.8$, respectively.
}
 \label{fig2a}
\end{figure}%

In contrast to this agreement, we also find statistical properties
 that are clearly different between the two systems.
The occupation time $T_+$,
 i.e., the length of time spent by the positive sign,
 is a quantity well-studied in two-state stochastic processes
 \cite{Lamperti-TAMS1958,Godreche.Luck-JSP2001}
 and in more general scale-invariant phenomena
 (see, e.g., \cite{Dornic.Godreche-JPA1998}).
It is simply related to the time-averaged sign
 $\bar\sigma\equiv(1/T)\int_0^T\sigma(x,t)\rd{}t$ by $\bar\sigma=2T_+/T-1$.
For RP with $0<\theta<1$, Lamperti \cite{Lamperti-TAMS1958}
 showed that it does not converge to the ensemble average,
 but remains stochastic even for $T\to\infty$,
 with its pdf derived exactly as follows
 \cite{Lamperti-TAMS1958,Godreche.Luck-JSP2001}:
\begin{equation}
 \text{pdf}(\bar\sigma) = \frac{(2\sin\pi\theta/\pi)(1-\bar\sigma^2)^{\theta-1}}{(1\!+\!\bar\sigma)^{2\theta} + (1\!-\!\bar\sigma)^{2\theta} + 2(\cos\pi\theta)(1\!-\!\bar\sigma^2)^\theta}.  \label{eq:Lamperti}
\end{equation}
This distributional behavior of the time-averaged sign
 is a clear evidence of WEB in RP.
The corresponding pdfs obtained at different $T$
 for the circular KPZ interfaces [\figref{fig2a}(a) symbols]
 indeed indicate an asymptotic broad distribution,
 demonstrating that $\bar\sigma$ remains stochastic and
 does not converge to the ensemble average $\expct{\sigma}=-0.021$
 (determined by the GUE Tracy-Widom distribution)
 shown by the gray vertial line in \figref{fig2a}(a).
This demonstrates that KPZ indeed exhibits WEB,
 at least for the circular case.
On the other hand, the found distribution is clearly different
 from the Lamperti's one for RP with $\theta=0.8$ (black dashed line).
We find instead
 a nontrivial distribution universal within the circular KPZ subclass,
 as supported by good agreement between experiments and simulations
 (symbols and turquoise solid line).

\begin{figure}[t!]
 \centering
 \includegraphics[width=.8\hsize,clip]{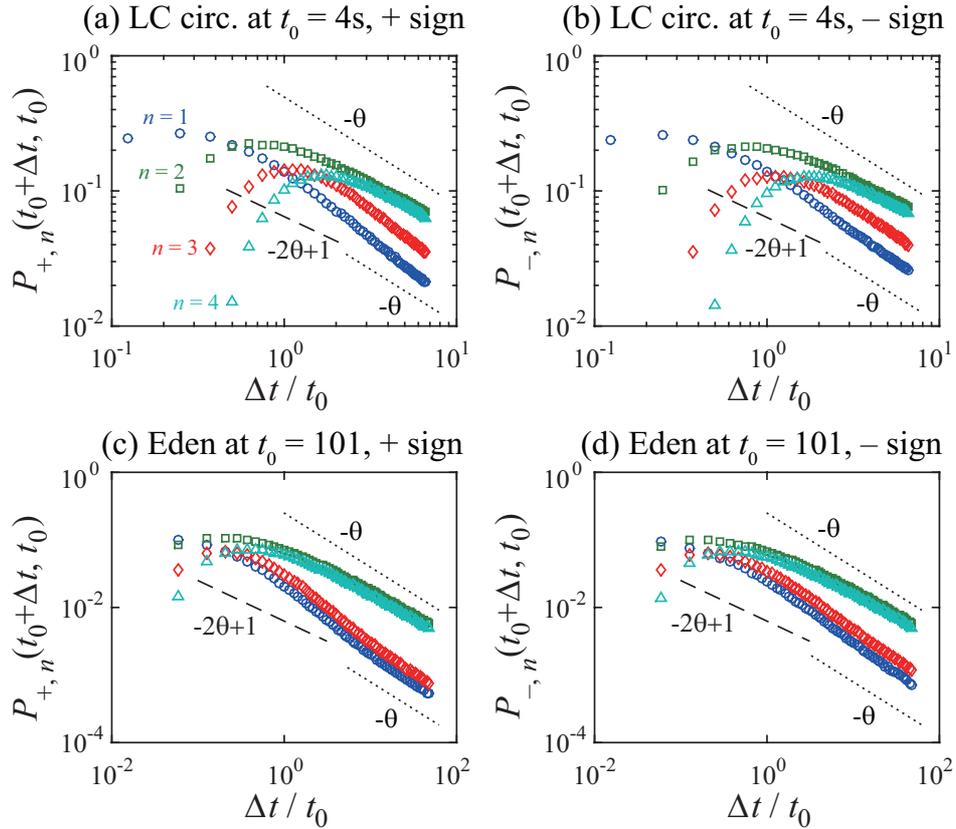}
 \caption{
Generalized persistence probabilities $P_{\pm,n}(t_0+\Delta t,t_0)$
 of the circular interfaces,
 measured for the positive (a,c) and negative (b,d) fluctuations
 (sign at $t_0$ is used)
 in the LC experiment (a,b) and the Eden model (c,d).
The dashed and dotted lines indicate exponents
 $-2\theta+1=-0.6$ and $-\theta=-0.8$, respectively.
The same set of colors/symbols and $n$ is used in all panels.
}
 \label{fig2b}
\end{figure}%

Another quantity of interest is the correlation function of sign,
 $C_{\rm{}sign}(t,t_0)\equiv\expct{\sigma(x,t)\sigma(x,t_0)}$.
This can be expanded by the generalized persistence probability
 $P_{\pm,n}(t,t_0)$, i.e.,
 the probability that the sign changes $n$ times between $t_0$ and $t$
 (hence $P_{\pm,0}(t,t_0)=P_\pm(t,t_0)$):
\begin{equation}
 C_{\rm{}sign}(t,t_0) = \sum_\pm P_\pm(t_0) \sum_{n=0}^\infty (-1)^n P_{\pm,n}(t,t_0),  \label{eq:SignCorrPers}
\end{equation}
 where $P_\pm(t_0)$ denotes the probability
 that fluctuations at $t_0$ take the sign indicated by the subscript.
For RP with $0<\theta<1$, 
 one can explicitly calculate the infinite sum of \eqref{eq:SignCorrPers}
 and obtain $C_{\rm{}sign}(t,t_0)\simeq{}\sum_\pm{}P_\pm(t_0)P_{\pm,0}(t,t_0)\sim{}(t/t_0)^{-\theta}$ \cite{Godreche.Luck-JSP2001}.
{In contrast, for the circular KPZ interfaces, we find}
 that $C_{\rm{}sign}(t,t_0)$
 decays as $C_{\rm{}sign}(t,t_0) \sim t^{-\bar\lambda}$
 with $\bar\lambda=1/3$ [\figref{fig2a}(b)] (see also footnote \ref{ft1}),
 in the same way as the rescaled correlation function $C_{\rm res}(t/t_0)$
 does [\eqref{eq:CorrFunc}].
Since the relation \pref{eq:SignCorrPers} holds generally
 and $P_{\pm,0}(t,t_0)=\int_{t-t_0}^\infty\text{pdf}(\tauf(t_0))\rd\tauf$
 is alike, the difference from RP
 should stem from $P_{\pm,n\geq1}(t,t_0)$,
 which encodes correlation between waiting times.
For RP, one can show
\begin{equation}
 P_{\pm,n\geq1}(t_0+\Delta{}t,t_0) \sim \begin{cases}
 \Delta{}t^{-2\theta+1} & \text{for $\Delta{}t\ll{}t_0$,} \\
 \Delta{}t^{-\theta} & \text{for $\Delta{}t\gg{}t_0$,}
 \end{cases}  \label{eq:GenPersRP}
\end{equation}
 with $t_0,\Delta{}t\gg\tau_0$ (see \ref{app:c}).
Now, for the circular KPZ interfaces,
 the results in \figref{fig2b} show that
 the long-time behavior seems to be consistent with that of RP\footnote{
\label{ft1}
For some quantities
 the asymptotic decay is only reached by the numerical data,
 obtained with longer time.
}, but the short-time behavior for odd $n$
 shows faster decay than that of RP
 (after the initial growth,
 which occurs at $\Delta t \lesssim n^{1/\theta}\tau_0$
 for RP; see \ref{app:c}). 
In other words, $P_{\pm,n}(t_0+\Delta{}t,t_0)$ has heavier weight
 in the short-time regime for odd $n$.
This difference from RP gives nontrivial contribution to the sum
 in \eqref{eq:SignCorrPers}, which is absent for RP.
We consider that this is how the different behavior
 of the correlation function arises,
 which captures, for KPZ, the characteristic time correlation
 of the (non-binarized) KPZ-class fluctuations.

\subsection{Flat interfaces}

\begin{figure}[t!]
 \centering
 \includegraphics[width=.8\hsize,clip]{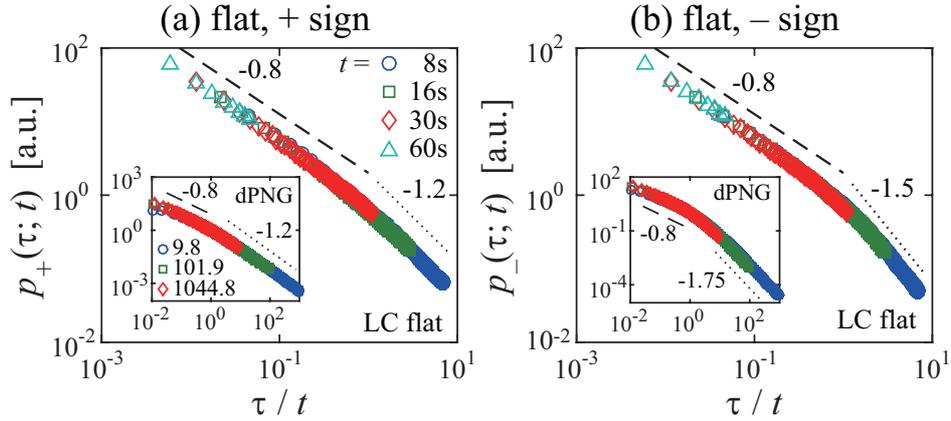}
 \caption{
Waiting-time distributions (ccdfs) $p_\pm(\tau;t)$ against $\tau/t$
 for the flat interfaces, obtained at different $t$
 in the LC experiment (main panels) and the dPNG model (insets).
The dashed and dotted lines are guides to the eyes,
 indicating exponents labeled alongside,
 though for the dotted lines
 we expect larger asymptotic exponents (see text).
The same set of colors/symbols and $t$ is used in both panels.
The ordinates are arbitrarily shifted.
}
 \label{fig3a}
\end{figure}%

Now we turn our attention to the flat interfaces.
Figure~\ref{fig3a} shows
 the waiting-time distribution (ccdf) $p_\pm(\tau;t)$
 for the flat LC experiment (main panels)
 and the dPNG model (insets).
At short waiting times,
 we identify power-law decay
 with exponent $-0.8$.
This exponent seems to be different from $-2/3$
 previously observed for a related quantity in the KPZ stationary state
 \cite{Kallabis.Krug-EL1999}\footnote{
The exponent measured in \cite{Kallabis.Krug-EL1999} was
 about the persistence probability of the sign
 of $\Delta h(x,t) - \Delta h(x,t_0)$
 with $\Delta h(x,t) = h(x,t) - \int_0^L h(x,t) \rd x/L$,
 whereas we measure here the waiting-time distribution,
 or the persistence probability
 of the sign of $\delta h(x,t) = h(x,t) - \expct{h(x,t)}$
 with the condition $\delta h(x,t_0)=0$.
Although both probabilities concern the first return to zero,
 the different definitions may lead to different exponent values.
},
 but is identical to the one
 we found for the circular interfaces (\figref{fig1a}).
For the flat interfaces (\figref{fig3a}), however, this power law
 is followed by another one with larger (in magnitude) exponent
 for longer waiting times,
 which now takes different values between positive and negative fluctuations.
The measured exponents do not seem to reach their asymptotic values
 within our observation time, increasing gradually with $\tau$,
 but they are clearly asymmetric with respect to the sign,
 in sharp contrast with the exponent for the shorter waiting times
 or for the circular interfaces.
Moreover, $p_\pm(\tau;t)$ taken at different $t$ overlaps
 when it is plotted against $\tau/t$ (\figref{fig3a}).
This indicates that the value of $\tau$
 separating the two power-law regimes is not constant, but grows with $t$,
 showing the aging property of the waiting-time distribution.

\begin{figure}[t!]
 \centering
 \includegraphics[width=.8\hsize,clip]{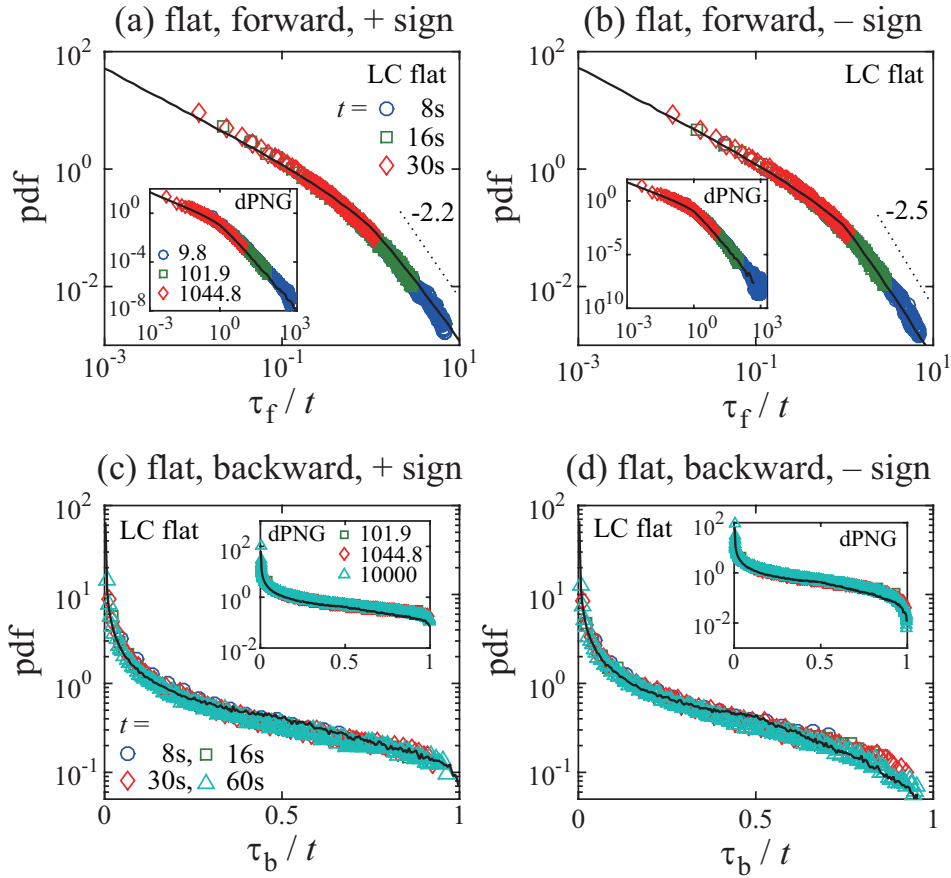}
 \caption{
Distributions (pdfs) of the forward (a,b) and backward (c,d) recurrence times,
 $\tauf$ and $\taub$, respectively,
 for the flat interfaces at different $t$
 in the LC experiment (main panels) and the dPNG model (insets).
$\tauf$ and $\taub$ are rescaled by $t$.
The black lines indicate numerical results for the 2-step RP at $t=10^7$.
The data are normalized so that
 they have the same statistical weight
 in the range covered by their abscissa.
The same colors/symbols correspond to the same $t$ in all panels.
}
 \label{fig3b}
\end{figure}%

These results on the flat-KPZ waiting-time distribution
 lead us to introduce a variant of RP with two power-law regimes,
 called hereafter the 2-step RP model
 \footnote{Strictly, since the evolution of the 2-step RP model
 [\eqref{eq:WaitingTimeRP2}] depends on $t$, it is \textit{not}
 in the scope of the models considered in the renewal theory.}:
\begin{equation}
 p_\pm(\tau;t) = 
 \begin{cases}
 \(\dfrac{\tau}{\tau_0}\)^{-\theta} & \text{for $\tau_0 \leq \tau \leq t$}, \\
 \(\dfrac{t}{\tau_0}\)^{-\theta} \(\dfrac{\tau}{t}\)^{-\theta'_\pm} & \text{for $t \leq \tau$}.
 \end{cases} \label{eq:WaitingTimeRP2}
\end{equation}
We then solved it numerically with $\theta=0.8,~\theta'_+=1.2,~\theta'_-=1.5$
 (as observed experimentally) 
in the following way: 
First, the initial sign was chosen to be either $+$ or $-$
 with the equal probability.
The first waiting time was generated according to
 $p_\pm(\tau;t)=(\tau/\tau_0)^{-\theta'_\pm}~(\tau\geq\tau_0)$,
 because \eqref{eq:WaitingTimeRP2} is invalid for $t=0$.
Subsequent waiting times were generated by \eqref{eq:WaitingTimeRP2},
 until the time (cumulative sum of waiting times)
 exceeds the recording time $T_{\rm{}tot}$.
We sampled $10^6$ independent realizations to investigate
 statistical properties of this 2-step RP model.

Now we compare this 2-step RP model and the flat KPZ interfaces.
Figure \ref{fig3b} shows the forward and backward
 recurrence-time distributions.
We find that these quantities for the flat KPZ interfaces
 are reproduced by the 2-step RP model reasonably well,
 similarly to those for the circular interfaces found in agreement
 with the standard RP.
Aging of the recurrence-time distributions is also clear
 in both cases.
Note however that, while $\text{pdf}(\tauf)\sim\tauf^{-\theta}$
 is known to hold for the standard RP [\eqref{eq:WaitingTimeRP}]
 with $1<\theta<2$ \cite{Godreche.Luck-JSP2001},
 our 2-step RP rather indicates $\text{pdf}(\tauf)\sim{}\tauf^{-\theta'_\pm-1}$
 [dotted lines in \figref{fig3b}(a,b)].
Since $\text{pdf}(\tauf)$ is given by
 the derivative of the persistence probability,
 this implies $\theta'_\pm=\theta_\pm^{\rm(p)}$, 
 hence asymptotically $\theta'_+=1.35$ and $\theta'_-=1.85$ are expected
 for the flat KPZ subclass.

\begin{figure}[t!]
 \centering
 \includegraphics[width=.8\hsize,clip]{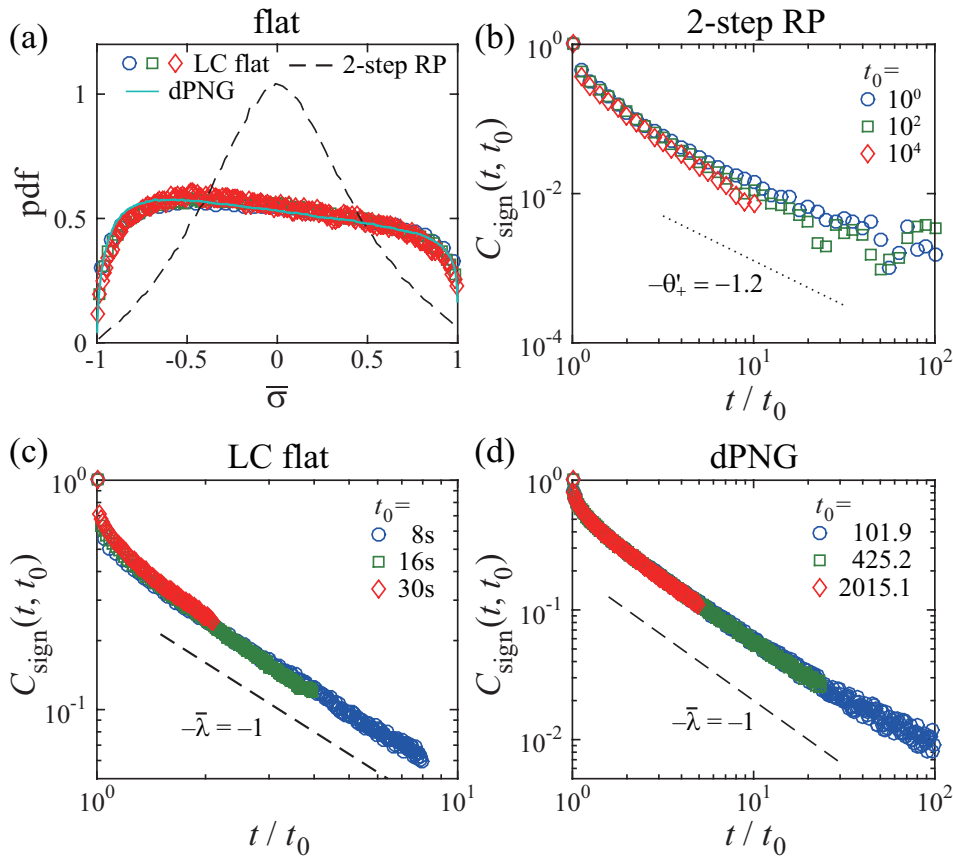}
 \caption{
Time-averaged sign distribution (a) and sign correlation function (b-d)
 for the flat interfaces (a,c,d) and the 2-step RP (b).
(a) Pdf of the time-averaged sign
 $\bar\sigma=(1/T)\int_0^T\sigma(x,t)\rd{}t$
 for the LC experiment ($T=15\unit{s},30\unit{s},63\unit{s}$
 for circles, squares, and diamonds, respectively) and
 the dPNG model ($T=10000$; turquoise line),
 compared to numerical data for the 2-step RP ($T=10^7$; dashed line).
The gray vertical line indicates the ensemble-averaged value
 $\expct{\sigma}=-0.0316$.
The existence of the broad asymptotic distribution is a direct evidence
 of WEB in the flat KPZ subclass,
 but in the form different from that of the circular case.
(b-d) Correlation function of sign,
 $C_{\rm{}sign}(t,t_0)=\expct{\sigma(x,t)\sigma(x,t_0)}$,
 at different $t_0$
 for the 2-step RP ($T_{\rm{}tot}=10^6$) (b),
 the LC flat interfaces (c), and the dPNG model (d).
The dashed lines in the panels (c,d) indicate the exponent $-\bar\lambda=-1$,
 while the dotted line in the panel (b) shows $-\theta'_+ = -1.2$.
}
 \label{fig4}
\end{figure}%

In contrast to this agreement in the recurrence-time distributions,
 the distribution of the time-averaged sign 
 $\bar\sigma=(1/T)\int_0^T\sigma(x,t)\rd{}t$ turns out to be
 different between the flat KPZ subclass and the 2-step RP [\figref{fig4}(a)],
 analogously to the results for the circular interfaces.
More specifically, both the flat KPZ subclass and the 2-step RP
 are found to show asymptotic broad distributions [\figref{fig4}(a)],
 hence both of them exhibit WEB, but the distributions are again
 clearly different between the two systems.
Note here that the time-averaged sign distribution for the standard RP
 [\eqref{eq:WaitingTimeRP}] with $\theta>1$ becomes
 infinitely narrow in the limit $t\to\infty$ \cite{Godreche.Luck-JSP2001};
 this is however not the case here, despite $\theta'_\pm>1$.
The existence of the broad distribution results from the aging
 of the waiting-time distribution, i.e., from the fact that
 the crossover time in the waiting-time distribution grows with $t$
 [see \figref{fig3a} and \eqref{eq:WaitingTimeRP2}].

\begin{figure}[t!]
 \centering
 \includegraphics[width=.8\hsize,clip]{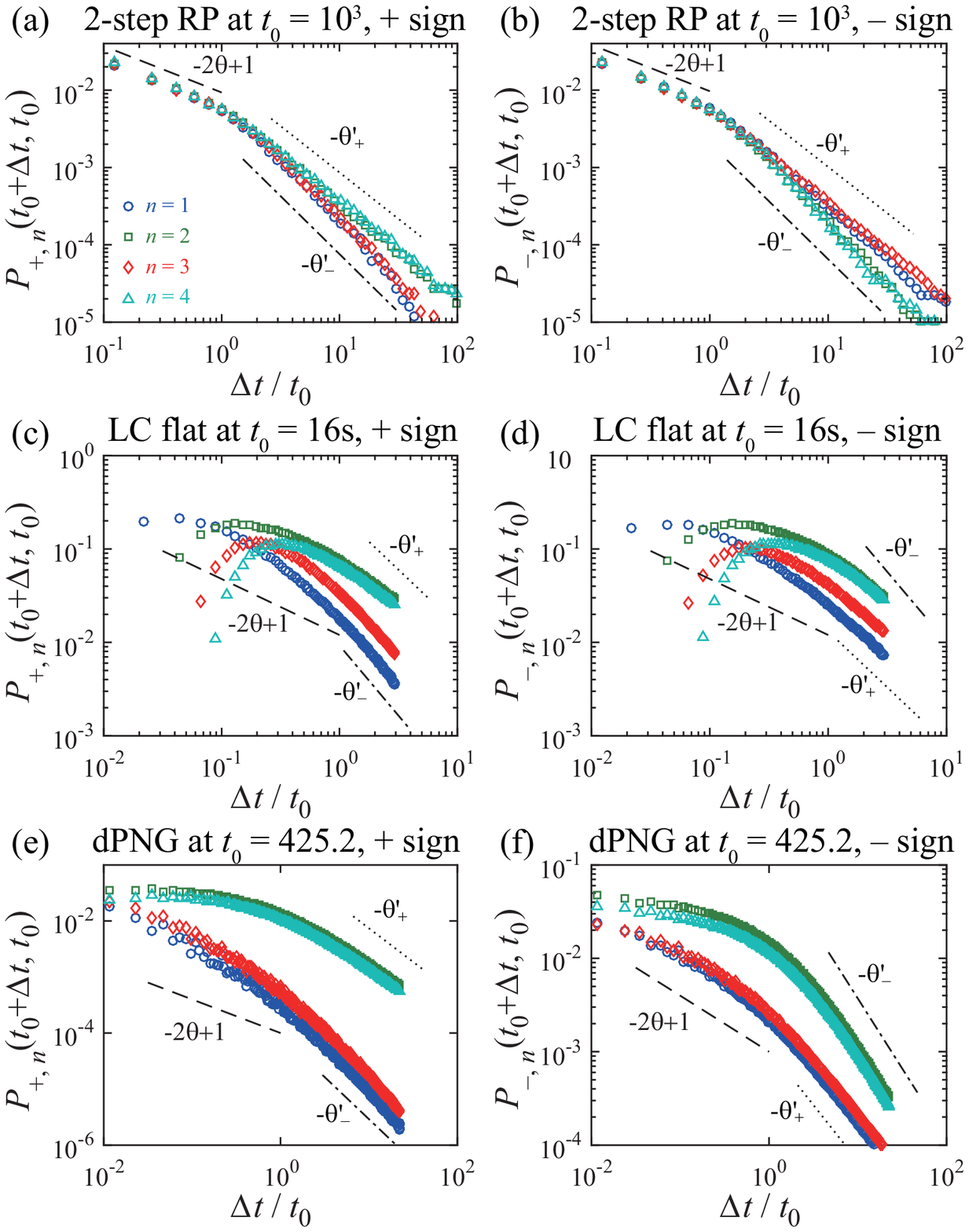}
 \caption{
Generalized persistence probabilities $P_{\pm,n}(t,t_0)$
 for the 2-step RP ($T_{\rm{}tot}=10^6$) (a,b)
 and for the flat KPZ-class interfaces
 [LC experiment (c,d) and dPNG model (e,f)],
 measured for the positive (a,c,e) and negative (b,d,f) signs.
The dashed lines in all panels indicate the exponent $-2\theta+1=-0.6$
 found in the short-time regime ($\Delta t \ll t_0$) of the 2-step RP.
The dotted and dot-dashed lines are guides for the eyes
 indicating exponents $-\theta'_+$ and $-\theta'_-$, respectively,
 which characterize the long-time regime for the 2-step RP
 [see \eqref{eq:GenPersRP2}].
The same set of colors/symbols and $n$ is used in all panels.
}
 \label{fig5}
\end{figure}%

The difference between the flat KPZ subclass and the 2-step RP
 is also detected in the correlation function of sign,
 $C_{\rm{}sign}(t,t_0)=\expct{\sigma(x,t)\sigma(x,t_0)}$:
 while our simulations of the 2-step RP show
 $C_{\rm{}sign}(t,t_0)\sim{}t^{-\theta'_+}$ [\figref{fig4}(b)],
 for the flat interfaces it decays as $t^{-\bar\lambda}$ with
 $\bar\lambda=1$ [\figref{fig4}(c,d)],
 the characteristic exponent
 for the decorrelation of the flat KPZ subclass
 [see \eqref{eq:CorrFunc}].
Similarly to the circular case,
 this difference results from correlation of waiting times,
 which can be characterized by the generalized persistence probability
 $P_{\pm,n}(t_0+\Delta{}t,t_0)$.
For the 2-step RP, i.e., in the absence of correlation,
 we numerically find [\figref{fig5}(a,b)]
\begin{equation}
 P_{\pm,n\geq1}(t_0+\Delta{}t,t_0) \sim \begin{cases}
 \Delta{}t^{-2\theta+1} & \text{for $\Delta{}t\ll{}t_0$,} \\
 \Delta{}t^{-\theta'_{\pm}} & \text{for $\Delta{}t\gg{}t_0$,}
 \end{cases}  \label{eq:GenPersRP2}
\end{equation}
 where in the latter case
 the two double signs are set to be the same sign for even $n$
 and the opposite ones for odd $n$.
This long-time behavior can also be seen in the flat KPZ subclass
 [\figref{fig5}(c,d) for the LC experiment and (e,f) for the dPNG model]. 
In contrast, short-time behavior of $P_{\pm,n\geq1}$ is found to be
 different between the 2-step RP and the flat KPZ subclass
 [compare data and the dashed lines in \figref{fig5}(c-f)],
 the latter carrying heavier weight in the short-time regime. 
Analogously to the circular case, such pronounced short-time behavior
 of $P_{\pm,n\geq1}$ seems to generate, via \eqref{eq:SignCorrPers},
 the characteristic decay of the correlation function $C_{\rm{}sign}(t,t_0)$
 slower than that of the 2-step RP [\figref{fig4}(b)].

\section{Concluding remarks}

We have shown an unexpected similarity
 between sign renewals of the KPZ-class fluctuations
 and RP, studied in the context of aging phenomena
 and WEB. 
Despite the fundamental difference between the two systems,
 we found, for the circular interfaces, that
 the KPZ waiting times obey simple power-law distributions
 identical to those defining RP,
 while those for the flat interfaces correspond
 to its straightforward extension with two power-law regimes
 [\eqref{eq:WaitingTimeRP2}, the 2-step RP model].
Further quantitative agreement has been found
 in the recurrence-time distributions
 (Figs.~\ref{fig1b} and \ref{fig3b}),
 from which the agreement in the persistence probability follows.
These quantities have remained theoretically intractable for KPZ,
 but now, following the agreement we found,
 their precise forms are revealed for the circular interfaces,
 thanks to the exact solutions for the original RP.
This also implies that recurrence-time statistics may be determined
 independently of the intercorrelation of waiting times, 
 contrary to the usual beliefs.

The correlated waiting times of KPZ otherwise generate characteristic
 aging properties of the KPZ-class fluctuations
 (\figsref{fig2a} and \ref{fig4}),
 especially their broad asymptotic distributions
 of the time-averaged sign.
This indicates WEB of the KPZ-class fluctuations,
 which turned out to be different from that of RP,
 and in fact also from other types of WEB,
 known from the studies of single-particle observations.
We therefore consider that the WEB found in this study
 is of a new kind, characteristic
 of many-body problems governed by the KPZ universality class.
This also implies that RP cannot be a proxy for the full KPZ dymamics;
 instead RP reproduces only some of the time-correlation properties of KPZ,
 surprisingly well, though.

In fact, such a partial similarity to RP was also argued in the past
 for the fractional Brownian motion (FBM),
 in the context of linear growth processes.
Krug et al. \cite{Krug.etal-PRE1997} showed,
 for the stationary state of linear growth processes,
 that the stochastic process $h(x,t) - h(x,t_0)$ is equivalent to FBM.
Its first-return time (corresponding to the waiting time of its sign)
 is then characterized by a power-law distribution 
 with exponent $\theta = 1-\beta$
 \cite{Hansen.etal-F1994,Ding.Yang-PRE1995,Krug.etal-PRE1997}
 with $\beta$ being the growth exponent, or the Hurst exponent of FBM.
Cakir \textit{et al.} \cite{Cakir.etal-PRE2006} then suggested
 that the sign of FBM would form RP,
 showing numerical observations of its persistence probability
 as a partial support,
 but it turned out later that the two models behave differently
 in other statistical quantities \cite{GarciaGarcia.etal-PRE2010},
 because of the intercorrelation of waiting times.
In our contribution,
 we studied the growth regime of nonlinear growth processes in the KPZ class
 and compared the sign of the stochastic process
 $h(x,t) - \expct{h(x,t)}$ with RP.
As already summarized, we showed thereby precise agreement
 in the waiting-time distribution and the persistence probability,
 but not in the other statistical properties we studied.
Understanding the mechanism of this partial agreement is an important issue
 left for future studies,
 all the more because no theoretical understanding has been made so far
 on persistence properties of the KPZ growth regime \cite{Bray.etal-AP2013}.
Such developments will also help to understand the deviations from RP,
 which we believe carry characteristic information
 of underlying growth processes
 (recall our results on the correlation function).
We hope this direction of analysis may afford a clue to elucidate
 hitherto unexplained time-correlation properties of the KPZ class.
We also believe that our approach may be useful to characterize
 other scale-invariant processes such as critical phenomena.

\begin{acknowledgements}
We acknowledge fruitful discussions
 with E.~Barkai, I.~Dornic, C.~Godr\`eche, and S.~N.~Majumdar.
This work is supported in part by KAKENHI from JSPS (No. JP25707033 and No. JP25103004), the JSPS Core-to-Core Program ``Non-equilibrium dynamics of soft matter and information'',
 and the National Science Foundation under Grant No. NSF PHY11-25915.
\end{acknowledgements}

\appendix
\def\thesection{Appendix \Alph{section}}

\section[\hspace{40pt}Studied systems]{Studied systems}  \label{app:a}

In this appendix we briefly describe
 the three systems studied in this paper,
 namely the LC experiment
 \cite{Takeuchi.etal-SR2011,Takeuchi.Sano-JSP2012},
 the off-lattice Eden model \cite{Takeuchi-JSM2012},
 and the dPNG model, all known to be in the KPZ class.
The experimental results are obtained
 from the raw data acquired
 in Refs.~\cite{Takeuchi.etal-SR2011,Takeuchi.Sano-JSP2012}.
The readers are referred to these publications
 for the complete description of the experimental system.

\subsection{LC experiment}

The experiment concerns fluctuating interfaces 
 between two turbulent regimes of electrically driven nematic liquid crystal,
 called the dynamic scattering modes 1 and 2 (DSM1 and DSM2, respectively)
 \cite{Takeuchi.Sano-PRL2010,Takeuchi.etal-SR2011,Takeuchi.Sano-JSP2012}.
The DSM1/DSM2 configuration
 can be argued to lie in pure two dimensions
\footnote{This is because the DSM2 state,
 known to consist of densely entangled topological defects,
 needs to break surface anchoring on the bottom and top plates
 to be sustained in the system.
},
 so the interfaces in between are one-dimensional.
Under sufficiently high applied voltage, here $26\unit{V}$,
 DSM2 is more stable than DSM1, and the interfaces grow
 until the whole system is occupied by DSM2.
The initial nucleus of the DSM2 state can be introduced by shooting
 laser pulses.
This allows us to study both circular and flat growing interfaces:
 circular interfaces grow from a point nucleus
 generated by focused laser pulses, while flat interfaces originate from
 a linear region of DSM2, created by linearly expanded laser pulses.

In Refs.~\cite{Takeuchi.etal-SR2011,Takeuchi.Sano-JSP2012},
 Takeuchi and Sano measured 955 circular interfaces
 over time length $30.5\unit{s}$
 and 1128 flat interfaces over $63\unit{s}$,
 and found the characteristic statistical properties
 of the circular and flat KPZ subclasses, respectively.
In the present study, we employ the same data sets
 which are guaranteed to belong to these subclasses,
 and analyze the sign of the height fluctuations as explained in the main text.
The sign renewals are detected at every $0.5\unit{s}$ and $0.35\unit{s}$
 for the circular and flat interfaces, respectively.

\subsection{Off-lattice Eden model}

Numerical data for circular interfaces
 are obtained with the off-lattice Eden model,
 the version introduced in Ref.~\cite{Takeuchi-JSM2012}
 which is sometimes called the off-lattice Eden D model.
While detailed descriptions can be found in Ref.~\cite{Takeuchi-JSM2012},
 in this model, one starts with a round particle of unit diameter
 placed at the origin of two-dimensional continuous space.
At each time step, one randomly chooses one of the $N$ existing particles,
 and attempts to put an identical particle next to it
 in a direction randomly chosen from the range $[0,2\pi)$.
If the new particle does not overlap any existing particles,
 it is added as attempted, otherwise the particle is discarded.
Time $t$ is then increased by $1/N$,
 whether the attempt is adopted or not.
Particles without enough adjacent space, to which no particle
 can be added any more, are labelled inactive and excluded
 from the particle counter $N$ (but can still block new particles).
Since we are interested in the interface,
 or specifically the outermost closed loop of adjacent particles,
 particles surrounded by the interface are also marked inactive and treated
 likewise.
This model was previously shown to belong to the circular KPZ subclass
 \cite{Takeuchi-JSM2012}.
The data presented in the present paper are newly obtained
 from $5000$ independent simulations of time length $5000$,
 and the sign renewals are detected at every time unit.

\subsection{dPNG model}

Numerical simulations of flat interfaces are performed
 with the dPNG model.
This is a discretized version of the PNG model,
 which is one of the exactly solvable models
 in the $(1+1)$-dimensional KPZ class \cite{Prahofer.Spohn-PRL2000}.
The evolution of the height variable $h(x = i \Delta x, t=n \Delta t)$
 of the dPNG model, with non-negative integers $h,i,n$,
 is given by the following equation:
\begin{equation}
 h(x,t+\Delta t) = \max\{ h(x-\Delta x, t), h(x,t), h(x+\Delta x,t) \} + \eta(x,t),
\end{equation}
 where $\eta(x,t)$ is
 an independent and identically distributed random variable
 generated from the geometric distribution,
 $\text{Prob}(\eta = k) = (1-p)^kp$ with $p = \rho \Delta x \Delta t$.
The original PNG model with nucleation rate $\rho$
 and nucleus expansion rate $\Delta x/\Delta t$ is retrieved
 by the continuum limit $\Delta x \to 0$ and $\Delta t \to 0$.

In our study, we set
 $\rho = 2, \Delta x = \Delta t = 0.1$ and the periodic boundary condition
 $h(L, t) = h(0,t)$ with $L = 10^4$ (or $10^5$ lattice units).
We start from the flat initial condition $h(x,0)=0$
 and evolve the system until $t = 10^4$
 by $10^4$ independent simulations.
The sign renewals are detected at every time step,
 i.e., $\Delta t = 0.1$ time unit.
Note that the dPNG model with $\Delta x = \Delta t = 0.1$
 shows the same universal statistical properties as the original PNG model,
 provided that the height variable $h(x,t)$ is appropriately rescaled
 using non-universal scaling coefficients
 [$v_\infty$ and $\Gamma$ in \eqref{eq:Height}].
The values of the scaling coefficients depend on $\Delta x$ and $\Delta t$:
 for example, they are estimated
 at $v_\infty \approx 2.2408$ and $\Gamma \approx 1.571$
 for the dPNG model studied here
 (the evaluation method
 described in Ref.~\cite{Takeuchi.Sano-JSP2012} is used),
 while the values for the original PNG model
 (corresponding to $\Delta x, \Delta t \to 0$) are
 $v_\infty = 2$ and $\Gamma = 1$.

\section[\hspace{40pt}Generalized persistence probability for RP]{Generalized persistence probability for RP}  \label{app:c}

Here we derive two asymptotic behaviors of the generalized persistence probability for the renewal process with a power-law 
waiting-time distribution. 
We assume Eq.~(\ref{eq:WaitingTimeRP}) with $\theta<1$
 for the waiting time distribution. 
Thus,  
the Laplace transform of the probability density function (pdf) of waiting times $\tau$, $\rho(\tau) \equiv p'(\tau)$, is given by
\begin{equation}
\hat{\rho}(s) = 1 - a s^\theta + \mathcal{O}(s),
\end{equation}
with $a=\Gamma(1-\theta) \tau_0^\theta$ \cite{Godreche.Luck-JSP2001}. 
The generalized persistent probability can be represented by
\begin{equation}
P_n (t_0 + \Delta t,t_0) 
 = \text{Prob}[\Delta t_n < \Delta t; t_0] - \text{Prob}[\Delta t_{n+1} < \Delta t;t_0],
\end{equation}
where $\text{Prob}[\Delta t_n<\Delta t; t_0]$ is the probability that 
 $\Delta t_n \equiv \tauf(t_0) + \tau_2 + \cdots + \tau_n < \Delta t$ holds,
 with waiting times $\tau_i$ and the forward recurrence time $\tauf(t_0)$
 (time elapsed from $t_0$ to the first renewal event since then).
For $n \geq 1$,
the double Laplace transform of $P_n(t_0 + \Delta t,t_0)$
 with respect to $t_0$ and $\Delta t$
 can be calculated as follows:
\begin{align}
\hat{P}_n (s,u) &\equiv \int_0^\infty \int_0^\infty P_n (t_0 + \Delta t,t_0) \re^{-st_0 - u\Delta t} \rd t_0 \rd\Delta t \nonumber\\
&= \frac{\hat{f}_E(u;s) \hat{\rho}(u)^{n-1}}{u} - \frac{\hat{f}_E(u;s) \hat{\rho}(u)^{n}}{u},
\end{align}
where $\hat{f}_E(u;s)$ is the double Laplace transform of
 $\text{pdf}(\tauf(t_0); t_0)$,
 given by \cite{Godreche.Luck-JSP2001}
\begin{align}
\hat{f}_E(u;s) &\equiv \int_0^\infty \int_0^\infty \text{pdf}(\tauf;t_0) \re^{-u\tauf-st_0} \rd\tauf \rd t_0 \nonumber\\
&= \frac{\hat{\rho}(u) - \hat{\rho}(s)}{s-u} \frac{1}{1-\hat{\rho}(s)}.
\label{double_Lap_fe}
\end{align}
Therefore,
\begin{equation}
\hat{P}_n (s,u) = \frac{\hat{\rho}(u) - \hat{\rho}(s)}{s-u} \frac{1}{1-\hat{\rho}(s)} \hat{\rho}(u)^{n-1} \frac{1-\hat{\rho}(u)}{u}.
\label{gpp_1}
\end{equation}

Now we consider the following two asymptotic limits. 
For $u \ll s \ll \tau_0^{-1}$ ($\tau_0 \ll t_0\ll \Delta t$),
 we obtain
\begin{equation}
\hat{P}_n (s,u) 
\simeq \frac{1}{s} au^{\theta -1},
\end{equation}
where we used approximation
 $\hat{\rho}(u)^{n-1} \simeq 1-a(n-1)u^\theta \simeq 1$. 
In other words, $u$ is so small that
 $a(n-1)u^\theta \ll 1$, i.e., 
$\Delta t \gg (n-1)^{1/\theta} \tau_0 \simeq n^{1/\theta}\tau_0$. 
Then the inverse Laplace transform yields
\begin{equation}
P_n (t_0 + \Delta t, t_0) \simeq \left(\frac{\Delta t}{\tau_0}\right)^{-\theta},
\label{gpp_case1_1}
\end{equation}
 for $\tau_0 \ll t_0 \ll \Delta t$ 
 and $\Delta t \gg n^{1/\theta} \tau_0$. 
We note that this asymptotic behavior does not depend on $t_0$ nor $n$. 
In contrast,
 for $s \ll u \ll \tau_0^{-1}$ ($\tau_0 \ll \Delta t\ll t_0$),
\begin{equation}
\hat{P}_n (s,u) 
\simeq \frac{1}{s^\theta} au^{2\theta -2},
\end{equation}
where we used the same approximation as in the previous case. 
The inverse Laplace transform then yields
\begin{equation}
P_n (t_0 + \Delta t, t_0) \simeq \frac{\Gamma(1-\theta)}{\Gamma(\theta) \Gamma(2-2\theta)}\(\frac{t_0}{\tau_0}\)^{\!\!-1+\theta}\!\!\(\frac{\Delta t}{\tau_0}\)^{\!\!-2\theta+1}\hspace{-20pt},\hspace{20pt}
\label{gpp_case1_2}
\end{equation}
for $n^{1/\theta}\tau_0 \ll \Delta t \ll t_0$.
This asymptotic behavior is independent of $n$
 but does depends on $t_0$.


\begin{thebibliography}{10}
\providecommand{\url}[1]{{#1}}
\providecommand{\urlprefix}{URL }
\expandafter\ifx\csname urlstyle\endcsname\relax
  \providecommand{\doi}[1]{DOI~\discretionary{}{}{}#1}\else
  \providecommand{\doi}{DOI~\discretionary{}{}{}\begingroup
  \urlstyle{rm}\Url}\fi

\bibitem{Akimoto.Miyaguchi-PRE2013}
Akimoto, T., Miyaguchi, T.: Distributional ergodicity in stored-energy-driven
  l\'evy flights.
\newblock Phys. Rev. E \textbf{87}, 062,134 (2013)

\bibitem{Barabasi.Stanley-Book1995}
Barab\'asi, A.L., Stanley, H.E.: Fractal Concepts in Surface Growth.
\newblock Cambridge Univ. Press, Cambridge (1995)

\bibitem{Bouchaud-JPIF1992}
Bouchaud, J.P.: Weak ergodicity breaking and aging in disordered systems.
\newblock J. Phys. I (France) \textbf{2}, 1705--1713 (1992)

\bibitem{Bray.etal-AP2013}
Bray, A.J., Majumdar, S.N., Schehr, G.: Persistence and first-passage
  properties in nonequilibrium systems.
\newblock Adv. Phys. \textbf{62}, 225--361 (2013)

\bibitem{Brokmann.etal-PRL2003}
Brokmann, X., Hermier, J.P., Messin, G., Desbiolles, P., Bouchaud, J.P., Dahan,
  M.: Statistical aging and nonergodicity in the fluorescence of single
  nanocrystals.
\newblock Phys. Rev. Lett. \textbf{90}, 120,601 (2003)

\bibitem{Cakir.etal-PRE2006}
Cakir, R., Grigolini, P., Krokhin, A.A.: Dynamical origin of memory and
  renewal.
\newblock Phys. Rev. E \textbf{74}, 021,108 (2006)

\bibitem{Carrasco.etal-NJP2014}
Carrasco, I.S.S., Takeuchi, K.A., Ferreira, S.C., Oliveira, T.J.: Interface
  fluctuations for deposition on enlarging flat substrates.
\newblock New J. Phys. \textbf{16}, 123,057 (2014)

\bibitem{Corwin-RMTA2012}
Corwin, I.: The kardar-parisi-zhang equation and universality class.
\newblock Random Matrices Theory Appl. \textbf{1}, 1130,001 (2012)

\bibitem{Cox-Book1962}
Cox, D.R.: Renewal Theory.
\newblock Methuen, London (1962)

\bibitem{Ding.Yang-PRE1995}
Ding, M., Yang, W.: Distribution of the first return time in fractional
  brownian motion and its application to the study of on-off intermittency.
\newblock Phys. Rev. E \textbf{52}, 207--213 (1995)

\bibitem{Dornic.Godreche-JPA1998}
Dornic, I., Godr\`eche, C.: Large deviations and nontrivial exponents in
  coarsening systems.
\newblock J. Phys. A \textbf{31}, 5413--5429 (1998)

\bibitem{Dotsenko-JSM2013}
Dotsenko, V.: Two-time free energy distribution function in (1+1) directed
  polymers.
\newblock J. Stat. Mech. \textbf{2013}, P06,017 (2013)

\bibitem{Dotsenko-a2015}
Dotsenko, V.: Two-time free energy distribution function in the kpz problem.
\newblock arXiv:1507.06135  (2015)

\bibitem{Dynkin-IANSSM1955}
Dynkin, E.B.: Some limit theorems for sums of independent random variables with
  infinite mathematical expectations.
\newblock Izv. Akad. Nauk SSSR Ser. Mat. \textbf{19}, 247--266 (1955).
\newblock Selected Translations Math. Stat. Prob \textbf{1}, 171--189 (1961).

\bibitem{Feller-Book1971}
Feller, W.: An Introduction to Probability Theory and Its Applications, vol.~1,
  3 edn.
\newblock Wiley, New York (1968)

\bibitem{Ferrari.Spohn-a2016}
Ferrari, P.L., Spohn, H.: On the current time correlations for one-dimensional
  exclusion processes.
\newblock arXiv:1602.00486  (2016)

\bibitem{GarciaGarcia.etal-PRE2010}
Garc\'{\i}a-Garc\'{\i}a, R., Rosso, A., Schehr, G.: Longest excursion of
  fractional brownian motion: Numerical evidence of non-markovian effects.
\newblock Phys. Rev. E \textbf{81}, 010,102 (2010)

\bibitem{Godreche.Luck-JSP2001}
Godr\`eche, C., Luck, J.M.: Statistics of the occupation time of renewal
  processes.
\newblock J. Stat. Phys. \textbf{104}, 489--524 (2001)

\bibitem{Hansen.etal-F1994}
Hansen, A., Eng\o{}y, T., M\aa{}l\o{}y, K.J.: Measuring hurst exponents with
  the first return method.
\newblock Fractals \textbf{02}(04), 527--533 (1994)

\bibitem{He.etal-PRL2008}
He, Y., Burov, S., Metzler, R., Barkai, E.: Random time-scale invariant
  diffusion and transport coefficients.
\newblock Phys. Rev. Lett. \textbf{101}, 058,101 (2008)

\bibitem{Herault.etal-JSP2015}
Herault, J., P\'etr\'elis, F., Fauve, S.: $1/f^\alpha$ low frequency
  fluctuations in turbulent flows.
\newblock J. Stat. Phys. \textbf{161}, 1379--1389 (2015)

\bibitem{Herault.etal-EL2015}
Herault, J., P\'etr\'elis, F., Fauve, S.: Experimental observation of $1/f$
  noise in quasi-bidimensional turbulent flows.
\newblock Europhys. Lett. \textbf{111}, 44,002 (2015)

\bibitem{Jeon.etal-PRL2011}
Jeon, J.H., Tejedor, V., Burov, S., Barkai, E., Selhuber-Unkel, C.,
  Berg-S\o{}rensen, K., Oddershede, L., Metzler, R.: \textit{In Vivo} anomalous
  diffusion and weak ergodicity breaking of lipid granules.
\newblock Phys. Rev. Lett. \textbf{106}, 048,103 (2011)

\bibitem{Johansson-a2015}
Johansson, K.: Two time distribution in brownian directed percolation.
\newblock arXiv:1502.00941  (2015)

\bibitem{Kallabis.Krug-EL1999}
Kallabis, H., Krug, J.: Persistence of kardar-parisi-zhang interfaces.
\newblock Europhys. Lett. \textbf{45}, 20--25 (1999)

\bibitem{Kardar.etal-PRL1986}
Kardar, M., Parisi, G., Zhang, Y.C.: Dynamic scaling of growing interfaces.
\newblock Phys. Rev. Lett. \textbf{56}, 889--892 (1986)

\bibitem{Kriecherbauer.Krug-JPA2010}
Kriecherbauer, T., Krug, J.: A pedestrian's view on interacting particle
  systems, kpz universality and random matrices.
\newblock J. Phys. A \textbf{43}, 403,001 (2010)

\bibitem{Krug.etal-PRE1997}
Krug, J., Kallabis, H., Majumdar, S.N., Cornell, S.J., Bray, A.J., Sire, C.:
  Persistence exponents for fluctuating interfaces.
\newblock Phys. Rev. E \textbf{56}, 2702--2712 (1997)

\bibitem{Kuno.etal-JCP2000}
Kuno, M., Fromm, D.P., Hamann, H.F., Gallagher, A., Nesbitt, D.J.:
  Nonexponential ``blinking'' kinetics of single cdse quantum dots: A universal
  power law behavior.
\newblock J. Chem. Phys. \textbf{112}, 3117 (2000)

\bibitem{Lamperti-TAMS1958}
Lamperti, J.: An occupation time theorem for a class of stochastic processes.
\newblock Trans. Amer. Math. Soc. \textbf{88}, 380--387 (1958)

\bibitem{Manzo.etal-PRX2015}
Manzo, C., Torreno-Pina, J.A., Massignan, P., Lapeyre, G.J., Lewenstein, M.,
  Garcia~Parajo, M.F.: Weak ergodicity breaking of receptor motion in living
  cells stemming from random diffusivity.
\newblock Phys. Rev. X \textbf{5}, 011,021 (2015)

\bibitem{Margolin.Barkai-PRL2005}
Margolin, G., Barkai, E.: Nonergodicity of blinking nanocrystals and other
  l\'evy-walk processes.
\newblock Phys. Rev. Lett. \textbf{94}, 080,601 (2005)

\bibitem{Margolin.Barkai-JSP2006}
Margolin, G., Barkai, E.: Nonergodicity of a time series obeying l\'evy
  statistics.
\newblock J. Stat. Phys. \textbf{122}, 137--167 (2006)

\bibitem{Metzler.etal-PCCP2014}
Metzler, R., Jeon, J.H., Cherstvy, A.G., Barkai, E.: Anomalous diffusion models
  and their properties: non-stationarity{,} non-ergodicity{,} and ageing at the
  centenary of single particle tracking.
\newblock Phys. Chem. Chem. Phys. \textbf{16}, 24,128--24,164 (2014)

\bibitem{Miyaguchi.Akimoto-PRE2011}
Miyaguchi, T., Akimoto, T.: Intrinsic randomness of transport coefficient in
  subdiffusion with static disorder.
\newblock Phys. Rev. E \textbf{83}, 031,926 (2011)

\bibitem{Miyaguchi.Akimoto-PRE2015}
Miyaguchi, T., Akimoto, T.: Anomalous diffusion in a quenched-trap model on
  fractal lattices.
\newblock Phys. Rev. E \textbf{91}, 010,102 (2015)

\bibitem{Prahofer.Spohn-PRL2000}
Pr\"ahofer, M., Spohn, H.: Universal distributions for growth processes in
  $1+1$ dimensions and random matrices.
\newblock Phys. Rev. Lett. \textbf{84}, 4882--4885 (2000)

\bibitem{Schulz.etal-PRL2013}
Schulz, J.H.P., Barkai, E., Metzler, R.: Aging effects and population splitting
  in single-particle trajectory averages.
\newblock Phys. Rev. Lett. \textbf{110}, 020,602 (2013)

\bibitem{Schulz.etal-PRX2014}
Schulz, J.H.P., Barkai, E., Metzler, R.: Aging renewal theory and application
  to random walks.
\newblock Phys. Rev. X \textbf{4}, 011,028 (2014)

\bibitem{Singha-JSM2005}
Singha, S.B.: Persistence of surface fluctuations in radially growing surfaces.
\newblock J. Stat. Mech. \textbf{2005}, P08,006 (2005)

\bibitem{Spohn-a2015}
Spohn, H.: Fluctuating hydrodynamics approach to equilibrium time correlations
  for anharmonic chains.
\newblock arXiv:1505.05987  (2015)

\bibitem{Stefani.etal-PT2009}
Stefani, F.D., Hoogenboom, J.P., Barkai, E.: Beyond quantum jumps: Blinking
  nanoscale light emitters.
\newblock Phys. Today \textbf{62}, 34 (2009)

\bibitem{Stefani.etal-NJP2005}
Stefani, F.D., Zhong, X., Knoll, W., Han, M., Kreiter, M.: Memory in
  quantum-dot photoluminescence blinking.
\newblock New J. Phys. \textbf{7}, 197 (2005)

\bibitem{Tabei.etal-PNAS2013}
Tabei, S.M.A., Burov, S., Kim, H.Y., Kuznetsov, A., Huynh, T., Jureller, J.,
  Philipson, L.H., Dinner, A.R., Scherer, N.F.: Intracellular transport of
  insulin granules is a subordinated random walk.
\newblock Proc. Natl. Acad. Sci. USA \textbf{110}(13), 4911--4916 (2013)

\bibitem{Takeuchi-JSM2012}
Takeuchi, K.A.: Statistics of circular interface fluctuations in an off-lattice
  eden model.
\newblock J. Stat. Mech. \textbf{2012}, P05,007 (2012)

\bibitem{Takeuchi-JSM2014}
Takeuchi, K.A.: Experimental approaches to universal out-of-equilibrium scaling
  laws: turbulent liquid crystal and other developments.
\newblock J. Stat. Mech. \textbf{2014}, P01,006 (2014)

\bibitem{Takeuchi.Sano-PRL2010}
Takeuchi, K.A., Sano, M.: Universal fluctuations of growing interfaces:
  Evidence in turbulent liquid crystals.
\newblock Phys. Rev. Lett. \textbf{104}, 230,601 (2010)

\bibitem{Takeuchi.Sano-JSP2012}
Takeuchi, K.A., Sano, M.: Evidence for geometry-dependent universal
  fluctuations of the kardar-parisi-zhang interfaces in liquid-crystal
  turbulence.
\newblock J. Stat. Phys. \textbf{147}, 853--890 (2012)

\bibitem{Takeuchi.etal-SR2011}
Takeuchi, K.A., Sano, M., Sasamoto, T., Spohn, H.: Growing interfaces uncover
  universal fluctuations behind scale invariance.
\newblock Sci. Rep. \textbf{1}, 34 (2011)

\bibitem{Weigel.etal-PNAS2011}
Weigel, A.V., Simon, B., Tamkun, M.M., Krapf, D.: Ergodic and nonergodic
  processes coexist in the plasma membrane as observed by single-molecule
  tracking.
\newblock Proc. Natl. Acad. Sci. USA \textbf{108}, 6438--6443 (2011)

\bibitem{Wong.etal-PRL2004}
Wong, I.Y., Gardel, M.L., Reichman, D.R., Weeks, E.R., Valentine, M.T., Bausch,
  A.R., Weitz, D.A.: Anomalous diffusion probes microstructure dynamics of
  entangled f-actin networks.
\newblock Phys. Rev. Lett. \textbf{92}, 178,101 (2004)

\bibitem{Yamamoto.etal-SR2015}
Yamamoto, E., Akimoto, T., Yasui, M., Yasuoka, K.: Origin of 1/f noise in
  hydration dynamics on lipid membrane surfaces.
\newblock Sci. Rep. \textbf{5}, 8876 (2015)

\bibitem{Yamamoto.etal-SR2015a}
Yamamoto, E., Kalli, A.C., Akimoto, T., Yasuoka, K., Sansom, M.S.P.: Anomalous
  dynamics of a lipid recognition protein on a membrane surface.
\newblock Sci. Rep. \textbf{5}, 18,245 (2015)

\end{thebibliography}

\end{document}